\documentclass[journal]{IEEEtran}
\usepackage{amsmath,amsfonts}
\usepackage{algorithmic}
\usepackage{algorithm}
\usepackage{array}
\usepackage[caption=false,font=footnotesize,labelfont=rm,textfont=rm]{subfig}
\usepackage{textcomp}
\usepackage{stfloats}
\usepackage{url}
\usepackage{verbatim}
\usepackage{graphicx}
\usepackage{mathrsfs}
\usepackage{cite}
\usepackage{multirow}
\begin{document}
\title{Residual Channel Boosts Contrastive Learning for Radio Frequency Fingerprint Identification}

\author{Rui Pan, Hui Chen, Guanxiong Shen, \emph{Member, IEEE}, Hongyang Chen, \emph{Senior  Member, IEEE}
	\thanks{This work was supported in part by the National Natural Science Foundation of China (61871092), and Sichuan Science and Technology Program (2022NSFSC0904).
}\thanks{R. Pan and H. Chen are with the School of Information and Communication Engineering, University of Electronic Science and Technology of China, Chengdu 611731, P. R. China (e-mail: 202221010931@std.uestc.edu.cn; huichen0929@uestc.edu.cn). G. Shen is with the School of Cyber Science and Engineering, Southeast University, Nanjing, China (email: gxshen@seu.edu.cn) H. Y. Chen is with Zhejiang Lab, China (email: dr.h.chen@ieee.org). Corresponding author: Hui Chen.}}

\maketitle
\begin{abstract}

In order to address the issue of limited data samples for the deployment of pre-trained models in unseen environments, this paper proposes a residual channel-based data augmentation strategy for Radio Frequency Fingerprint Identification (RFFI), coupled with a lightweight SimSiam contrastive learning framework. By applying least square (LS) and minimum mean square error (MMSE) channel estimations followed by equalization, signals with different residual channel effects are generated. These residual channels enable the model to learn more effective representations. Then the pre-trained model is fine-tuned with 1\% samples in a novel environment for RFFI. Experimental results demonstrate that our method significantly enhances both feature extraction ability and generalization while requiring fewer samples and less time, making it suitable for practical wireless security applications.
\end{abstract}


\section{Introduction}
\IEEEPARstart{W}{ITH} the rapid development of the Internet of Things (IoT) devices, secure authentication has become increasingly critical. Traditional encryption-based methods impose significant overhead on low-cost, energy-constrained IoT devices~\cite{zhang2020a}. Recently, radio frequency fingerprint identification (RFFI) has been proposed to enhance the security of these low-cost IoT devices. In a nutshell, it extracts unique hardware impairments as device identifiers, which are generated during the manufacturing process and are difficult to counterfeit~\cite{zhang2021radio}.
Conventional RFFI needs manual feature extraction with expertise and yields unsatisfactory results. Deep learning (DL) is more efficient that it enables the automatic extraction of features, bypassing the limitations of manual approaches. Features such as IQ imbalance can be effectively extracted using DL methods, with convolutional neural network (CNN) demonstrating particular efficacy~\cite{restuccia2019deepradioid,al2020exposing,sankhe20no,jian2020deep,zhang2021radio,shen2022towards,Huang2022deep,hao2023contrastive, xing2023design,zha2023cross,tang2024causal,pan2024equalization,shen2024federated,liu2024overcoming}.

The transmission through wireless channels introduces variations in the data distribution between the training and test domains, resulting in performance degradation in cross-scenario applications. This dataset shift is especially prominent in RFFI due to dynamic environmental changes, which render pre-trained models ineffective. Moreover, the difficulty in collecting sufficient labeled data for retraining models in new environments further impedes real-world applications. To address this challenge, several studies propose extracting channel-independent features~\cite{shen2022towards, xing2023design, he24SQCE, tang2024causal}. While these methods effectively reduce the impact of channel variations and can be considered domain generalization approaches, domain generalization struggles to cover all possible domains~\cite{zhang2023adanpc}.

Achieving robust performance with few labeled data is valuable for real-world applications. To address the challenge of the limited label in RFFI, contrastive learning (CL) is commonly used for unsupervised pre-training~\cite{zha2023cross, hao2023contrastive, shen2024federated, liu2024overcoming}. CL improves feature extraction by maximizing similarity within positive pairs, and the pre-trained model is fine-tuned with a small labeled dataset for downstream tasks. While CL typically requires stronger data augmentation~\cite{chen2020simple}, existing methods often use computer vision-based techniques such as mask~\cite{zha2023cross}, flipping~\cite{hao2023contrastive,liu2024overcoming}, and rotation~\cite{liu2024overcoming}, which may not be well-suited for wireless communications. Some methods use simulated channels and noise~\cite{shen2024federated}, but these do not fully capture real-world conditions.

Equalization is also a viable way to deal with wireless channels and an essential process for data demodulation. In our previous work~\cite{pan2024equalization}, we employed equalization and domain adaptation to improve model performance in new environments. In this paper, we propose a data augmentation method for unsupervised CL based on equalization. Our main contributions are as follows:
\begin{itemize}[]
    \item We use different channel estimations to equalize raw signals, enabling the model to extract features from diverse residual channels in an unsupervised way and learn more meaningful representations.
    \item We propose a lightweight CL framework for RFFI, which significantly reduces memory requirements and computing resources.
    \item We evaluate our method on a simulated dataset and demonstrate that it reduces training time and labeled sample requirements while achieving competitive performance with pre-trained models.
\end{itemize}

The rest of the paper is organized as follows. Section II discusses the impact of IQ imbalance on devices and the role of channel equalization. Section III explores contrastive learning and fine-tuning. Section IV presents the experimental settings and results. Finally, conclusions are provided in Section V.

\section{IQ imbalance and residual channel}
The modulated signal is typically represented as a complex value and transmitted through the in-phase (I) and quadrature-phase (Q) branches, expressed as:
\begin{equation}
    \label{pure}
    x=x_I+jx_Q,
\end{equation}
where $x_I$ and $x_Q$ denote the I and Q components, respectively. Before transmission, the signal is upconverted to the RF band, and IQ imbalance will be brought to I and Q branches to the baseband signal \(x_{BB}\) as~\cite{zhang2021radio}:
\begin{equation}
\label{baseband}
\begin{aligned}
    x_{BB}&=(g_I\cos\theta x_I - g_Q\sin\theta x_Q) \\
    &+ j(g_I\sin\theta x_I + g_Q\cos\theta x_Q),
\end{aligned}
\end{equation}
where
\begin{equation}
    g_I=10^{0.5\frac{A}{20}},
\end{equation}
\begin{equation}
    g_Q=10^{-0.5\frac{A}{20}},
\end{equation}
\begin{equation}
    \theta = 0.5P\frac{\pi}{180}.
\end{equation}
\(A\) is the amplitude imbalance in dB and \(P\) is the phase imbalance in degree. Compared to the signal components, IQ imbalance tends to remain more stable over time, making it a distinct and consistent characteristic for CNN.

After transmission and carrier frequency offset compensation, the received signal in the frequency domain is 
\begin{equation}
    \label{received}
    y=hx_{BB}+n,
\end{equation}
where \(h\) is the channel and \(n\) is the corresponding noise.

To extract the transmitted signal, the channel is estimated by pilots, and equalization is applied to \(y\). There are two classical channel estimation methods, least squares (LS) estimation and minimum mean square error (MMSE) estimation~\cite{cho2010mimo, ye2018power}. LS estimation is effective without prior knowledge of the channel and noise. The estimated channel using LS is given by:
\begin{equation}
    \label{lsestimation}
    \hat{h}_{LS}=\frac{y_p}{x_p},
\end{equation}
where \(y_p\) and \(x_p\) are the received and transmitted pilot signals, respectively. However, LS estimation neglects the effect of noise, and its corresponding mean square error (MSE) increases with noise. Specifically, the MSE of LS estimation is inversely proportional to the signal-to-noise ratio (SNR), given by \(MSE(\hat{h}_{LS})=\frac{1}{SNR}\)~\cite{cho2010mimo}. To mitigate the impact of noise, MMSE estimation introduces a weight matrix applied to the LS estimation:
\begin{equation}
    \label{mmseestimation}
    \hat{h}_{MMSE}=\mathbf{R_{h\hat{h}_{LS}}}(\mathbf{R_{hh}}+\frac{1}{SNR}\mathbf{I})^{-1}\hat{h}_{LS},
\end{equation}
where \(\mathbf{R_{h\hat{h}_{LS}}}\) is the covariance matrix between the true channel \(h\) and \(\hat{h}_{LS}\), \(\mathbf{R_{hh}}\) is the autocovariance matrix of the true channel, \(\mathbf{I}\) is the identity matrix, and \((\cdot)^{-1}\) denotes the inverse of a matrix. By leveraging the statistical properties of the true channel, MMSE estimation effectively suppresses noise and enhances channel estimation accuracy. 

Then, equalization is applied to the received signals as follows:
\begin{equation}
    \label{equalization}
    \begin{aligned}
    \hat{x}_{BB}&=\frac{y}{\hat{h}} \\
    &=\frac{(\hat{h}+h-\hat{h})x_{BB}+n}{\hat{h}} \\
    &=x_{BB}+\Delta h,
    \end{aligned}
\end{equation}
where \(\Delta h\) represents the residual channel in the equalized signal. This \(\Delta h\) is the primary reason for performance degradation in cross-scenario applications, even when equalization is applied. Although MMSE equalization can further suppress the impact of noise, \(\Delta h\) provides an opportunity for data augmentation. It can be leveraged in CL to extract robust RFFI features.

\section{contrastive Learning for RFFI}
SimSiam~\cite{chen2021simsiam} employs a Siamese network architecture, where two identical CNNs are used to extract features from the input data, which can significantly reduce memory requirements, as only one network needs to be stored during training. Furthermore, SimSiam operates effectively with a normal batch size and only positive pairs, minimizing the demand for computing resources. In our work, equalized signals with I and Q branches are used as input data, which are efficiently processed using a lightweight 1D-CNN. These characteristics and the following experimental results demonstrate our method is a promising choice for RFFI.

\subsection{Unsupervised Training}
Data augmentation plays a critical role in CL, as it facilitates the learning of more robust and meaningful representations. In addition to utilizing different channel estimation, we apply additive white Gaussian noise (AWGN) augmentation to enhance the model's robustness and block-wise masking augmentation to encourage the learning of semantic-independent features~\cite{Huang2022deep, zha2023cross}.

\begin{figure*}
    \centering
    \includegraphics[width=7in]{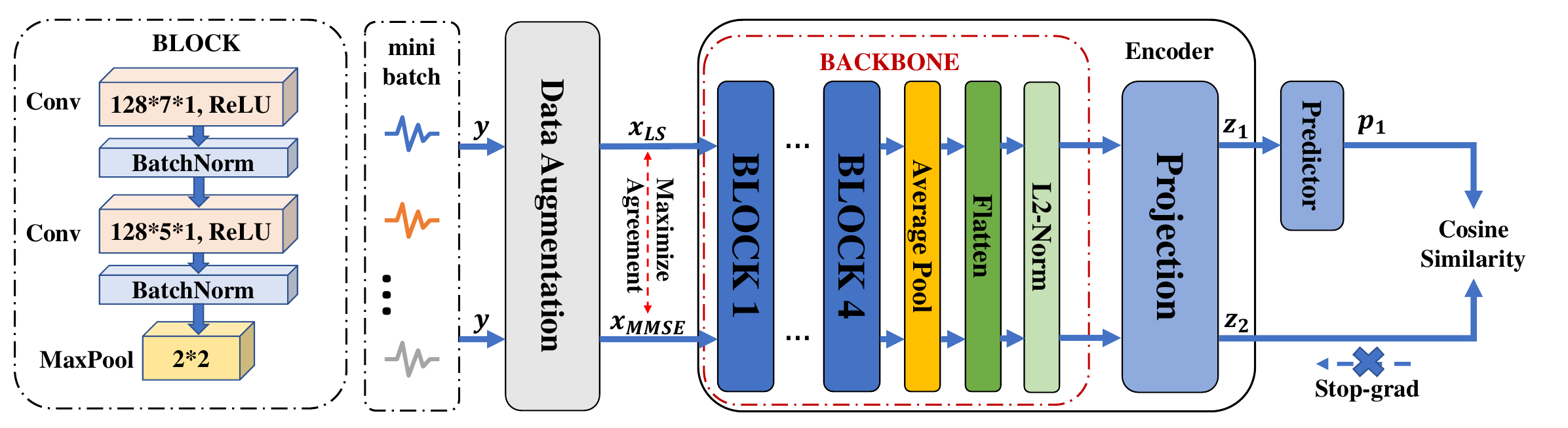}
    \caption{We employ a lightweight CNN, modified from our previous work~\cite{pan2024equalization}, as the backbone model for SimSiam. The backbone has a parameter size of only 2.58 MB, making it well-suited for IoT deployment. The encoder consists of four blocks, an adaptive pooling layer, a flattened layer, and an L2 normalization layer, which collectively enhance feature extraction and representation learning. The settings for the projection MLP and prediction MLP follow the SimSiam study. Input data in a small mini-batch undergoes various data augmentation. The agreement of equalized signals \(x_{LS}\) and \(x_{MMSE}\) are expected to be maximized by recovering the same \(x_{BB}\) in the present of \(\Delta h\).}
    \label{flow}
\end{figure*}

First, AWGN is added to the signals, followed by equalization using LS and MMSE methods under the corresponding SNR. Next, block-wise masking is applied to the augmented signals. After completing the data augmentation process, we maximize the cosine similarity between the two features extracted by the encoder. The process flow is illustrated in Fig. \ref{flow}. 

The SimSiam framework consists of a backbone model (e.g., a CNN) combined with a projection multilayer perceptron (MLP) serving as the encoder \(f(\cdot)\), a prediction MLP \(h(\cdot)\). Assume the data augmentation techniques for positive pairs are \(\tau_1\sim T_1\) and \(\tau_2\sim T_2\). The positive pairs come from the same inputs \(y\) as \(x_{LS}=\tau_1(y)\) and \(x_{MMSE}=\tau_2(y)\), which returns different equalized signals and other augmentations. Let the two outputs as \(p_1\stackrel{\triangle}{=}h(f(x_{LS}))\) and \(z_2\stackrel{\triangle}{=}f(x_{MMSE})\), their negative cosine similarity is defined as: 
\begin{equation}
    \label{cosloss}
    D(p_1,z_2)=-\frac{p_1}{||p_1||_2}\cdot\frac{z_2}{||z_2||_2},
\end{equation}
where \(||\cdot||_2\) is L2-norm. We optimize the model with symmetrized loss and use stop-gradient \(sg(\cdot)\) operation as 
\begin{equation}
    \begin{aligned}
    \label{symmetrized loss}
    \mathcal{L}_{CL}&=\frac{1}{2}D(p_1,sg(z_2)) + \frac{1}{2}D(p_2,sg(z_1)).
    \end{aligned}
\end{equation}
Stop-gradient operation is vital for SimSiam to work, the encoder on \(x_1\) receives only gradient from \(p_1\) in the first term but not \(z_1\) from the second term.

\subsection{Fine-Tuning}
After CL training, the backbone is retained, while the other parts of the architecture are discarded. It acquires stronger feature extraction capabilities but still requires fine-tuning to adapt to downstream tasks, such as RFFI. Compared to traditional supervised training, a powerful pre-trained model needs a few labeled datasets to achieve the supervised performance.

\begin{figure}
    \centering
    \includegraphics[width=3.2in]{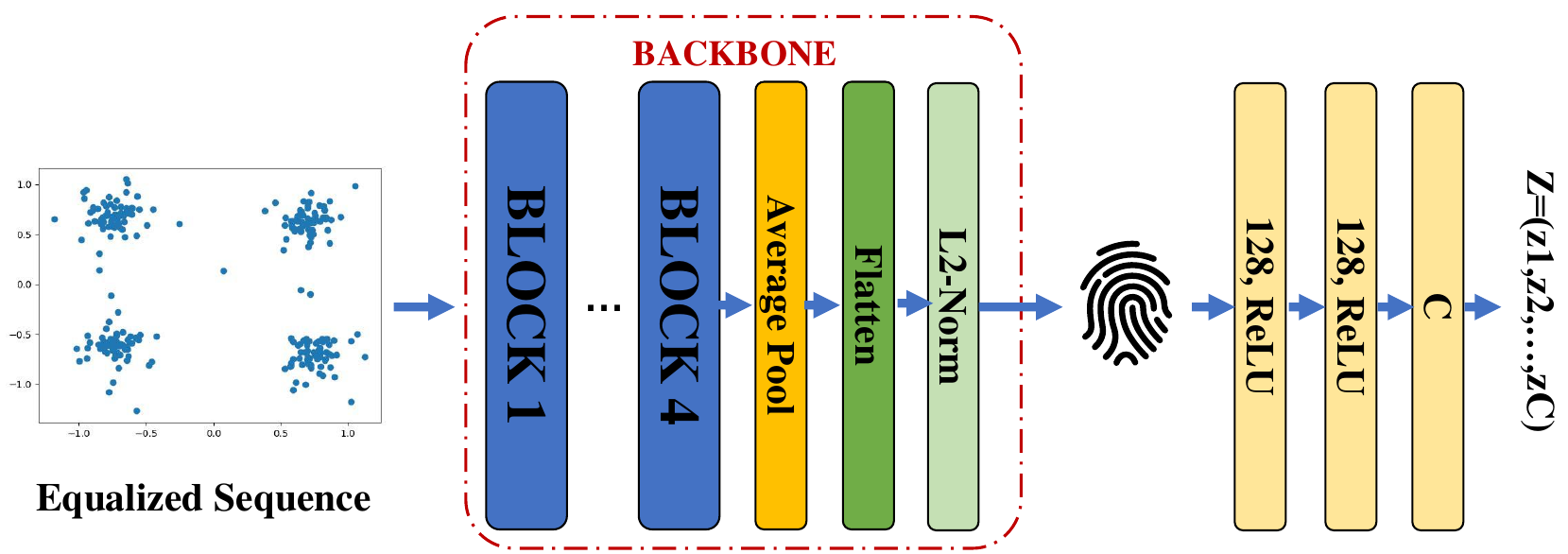}
    \caption{Equalized sequences are fed into the pre-trained backbone model, which extracts features for RFFI features. The classification head is trained using the labeled dataset to fine-tune the model for downstream tasks.}
    \label{fine-tune-network}
\end{figure}

The supervised fine-tuning is carried out in the standard way. First, the pre-trained backbone is used as a feature extractor. Then, a simple MLP, initialized randomly, is added to the backbone to form a standard CNN structure as shown in Fig. \ref{fine-tune-network}. The model is trained using Cross-Entropy (CE) loss, which backpropagates the gradient to update the model's parameters. The CE loss is defined as:
\begin{equation}
    \label{celoss}
    \mathcal{L}_{CE}=-\sum_{i=1}^{C}y_ilog(\sigma(z)_i),
\end{equation}
where \(y_i\) represents the label, and \(\sigma(\cdot)\) is a softmax operation for the final output of MLP, \(\boldsymbol{z}=(z_1,z_2,...,z_C)\) with C classes. The softmax operation is defined as:
\begin{equation}
    \label{softmax}
    \sigma(z)_i=\frac{e^{z_i}}{\sum_{j=1}^{C}e^{z_j}}.
\end{equation}

The details of our proposed method are summarized in Algorithm \ref{algo}. 
\begin{algorithm}[htbp]
    \caption{Contrastive Learning with Residual Channel}\label{algo}
    \begin{algorithmic}
    \STATE \textbf{Input:} Dataloader $D$, real channel $h$, SNR range, pilot $x_p$, epochs, encoder $f(\cdot)$, prediction MLP $h(\cdot)$
    \STATE \textbf{Output:} pre-trained backbone CNN
    \STATE \textbf{for} $epoch$ \textbf{in} epochs:
    \STATE \hspace{0.25cm} \textbf{for} $y$ \textbf{in} $D$:
    \STATE \hspace{0.5cm} \# \textbf{data augmentation}
    \STATE \hspace{0.5cm} $snr1$, $snr2$ = torch.randint(SNR range)
    \STATE \hspace{0.5cm} $y_1$, $y_2$ = AWGN($y$, $snr1$), AWGN($y$, $snr2$)
    \STATE \hspace{0.5cm} calculate $\hat{h}_{LS}$ by Eq. (\ref{lsestimation}) and $\hat{h}_{MMSE}$ by Eq.
    (\ref{mmseestimation})
    \STATE \hspace{0.5cm} $x_{LS}$, $x_{MMSE}$ = $\frac{y_1}{\hat{h}_{LS}}$, $\frac{y_2}{\hat{h}_{MMSE}}$
    \STATE \hspace{0.5cm} $x_{LS}$, $x_{MMSE}$ = mask($x_{LS}$), mask($x_{MMSE}$)
    \STATE \hspace{0.5cm} \# \textbf{contrastive learning}
    \STATE \hspace{0.5cm} $z_1$, $z_2$ = $f(x_{LS})$, $f(x_{MMSE})$
    \STATE \hspace{0.5cm} $p_1$, $p_2$ = $h(z_1)$, $h(z_2)$
    \STATE \hspace{0.5cm} calculate $\mathcal{L}_{CL}$ by Eq. (\ref{cosloss}) and Eq. (\ref{symmetrized loss})
    \STATE \hspace{0.5cm} backward $\mathcal{L}_{CL}$
    \STATE \hspace{0.5cm} update $f(\cdot)$, $h(\cdot)$
    \STATE \# \textbf{save model}
    \STATE Keep the backbone CNN and discard the rest
    
    \end{algorithmic}
\end{algorithm}

\section{Experimental Results}
\subsection{Experimental Settings}
We follow previous work's settings for IQ imbalance~\cite{pan2024equalization}. Using Python, we generate raw signals with IQ imbalance parameters as shown in Table \ref{iqimbal} for 7 devices. The imbalance parameters are evenly distributed in the range [-0.9, 0.9] and [-3, 3]. 
The FFT length is set to 64, with 52 valid subcarriers, and QPSK modulation is applied. The simulated data also includes a long training sequence (LTS) for channel estimation. We adopt the tapped delay line (TDL) model for the simulated channel~\cite{shen2024federated}, with an exponential power delay profile (PDP) to generate multipath power and the Jakes model to account for Doppler effects with the Doppler frequency randomly distributed uniformly in the range [0, 5] Hz. The RMS delay is set to 30~ns. The basic SNR is set to 20 dB for the TDL channel. Parameters refer to 3GPP specification \footnote{3GPP TR 38.901. “Study on channel model for frequencies from 0.5 to 100 GHz.” 3rd Generation Partnership Project}. 

\begin{table}[!t]
    \caption{IQ Imbalance Parameters Set for Devices}
    \label{iqimbal}
    \centering
    \begin{tabular}{c|c|c}
    \hline
    \textbf{Property}  & \textbf{Imbalance Range} & \textbf{Classes}    \\ \hline
    \textbf{A (dB)}     & [-0.9, 0.9]    & \multirow{2}{*}{7 Devices}                  \\
    \textbf{P (degree)} & [-3, 3]        &                                \\ \hline
    \end{tabular}
\end{table}

We conducted comparison experiments and verified that contrastive learning using different residual channels performs significantly better than using a single-channel estimation method. We transmitted 1000 packets for each device. The dataset consists of five data frames, and the preprocessed data is segmented into batches of size \(batches\times 2\times 260\). The CL and CNN frame is implemented using PyTorch with Adam~\cite{kingma2014adam} as the optimizer. We record the average loss and accuracy to show the model's performance. The system is evaluated using a PC equipped with an Intel i5-12600KF CPU, 16GB of RAM, and an NVIDIA RTX 3070 GPU.

\begin{figure*}
    \centering
    \subfloat[LS-only Features, $\overline{NMI}$=0.214]{\includegraphics[width=2.2in]{figures/ls\_feaspace.pdf}}
    \subfloat[Mixed Features, $\overline{NMI}$=0.504]{\includegraphics[width=2.2in]{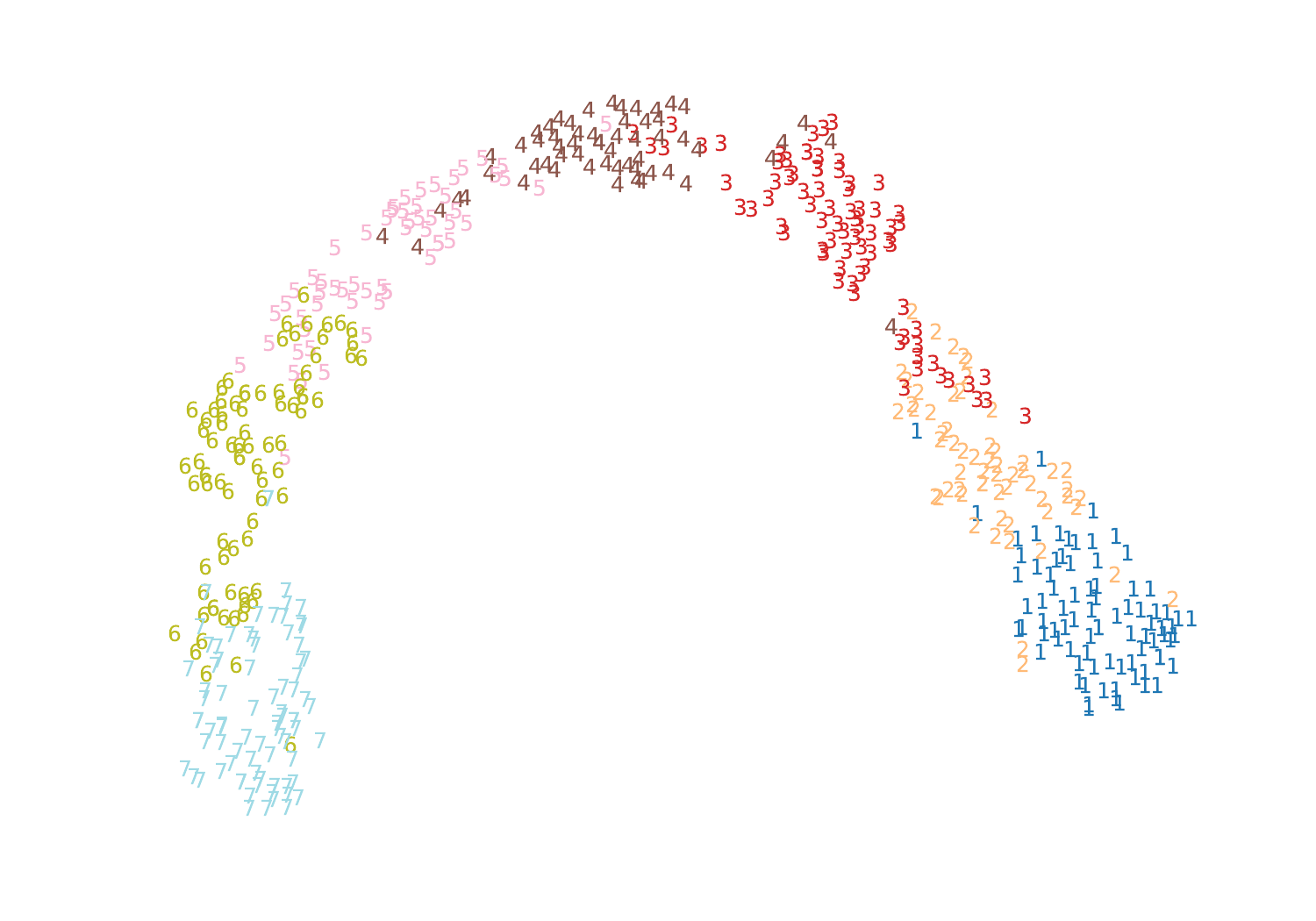}}
    \subfloat[MMSE-only Features, $\overline{NMI}$=0.323]{\includegraphics[width=2.2in]{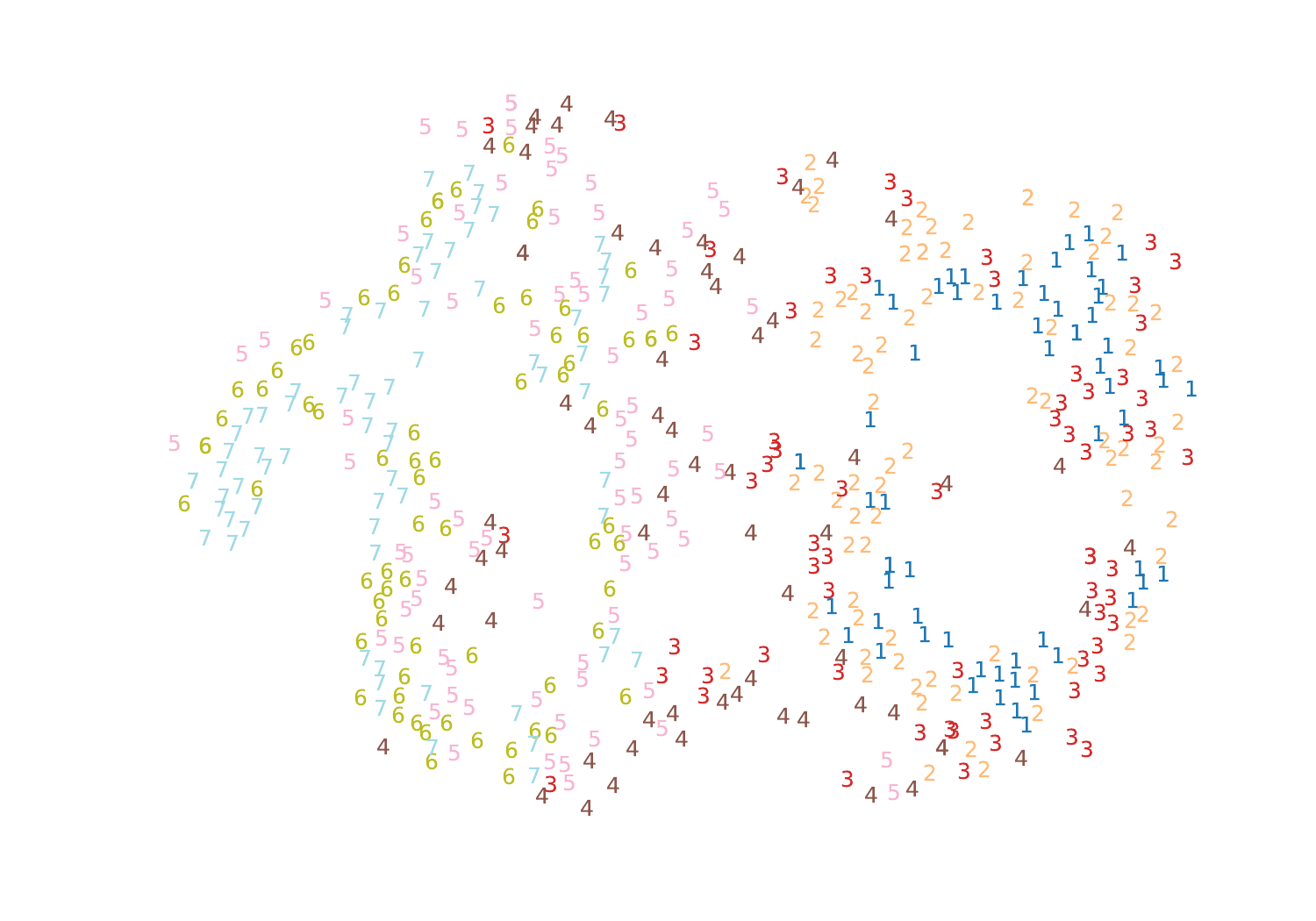}} \\
    \subfloat[LS-only loss, $\overline{time}$=178s]{\includegraphics[width=2.2in]{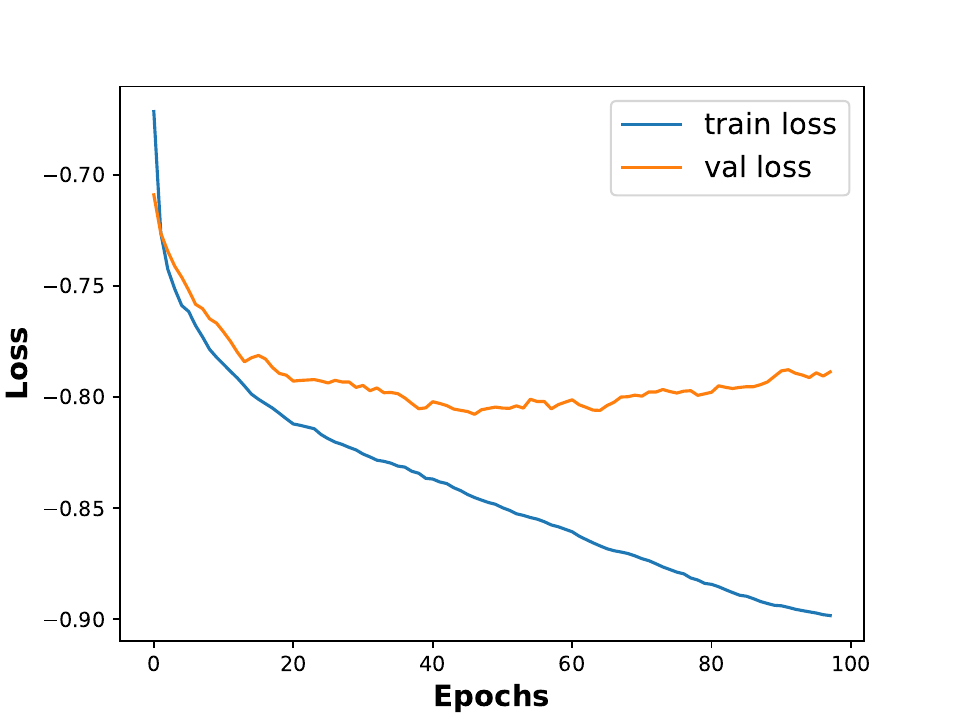}}
    \subfloat[Mixed loss, $\overline{time}$=319s]{\includegraphics[width=2.2in]{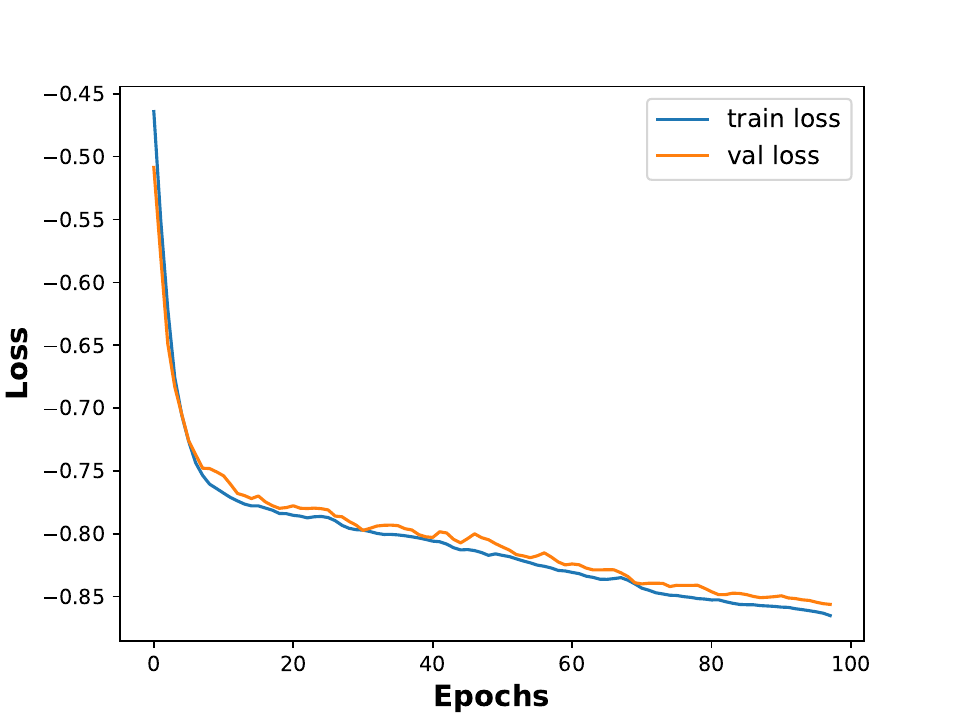}}
    \subfloat[MMSE-only loss, $\overline{time}$=446s]{\includegraphics[width=2.2in]{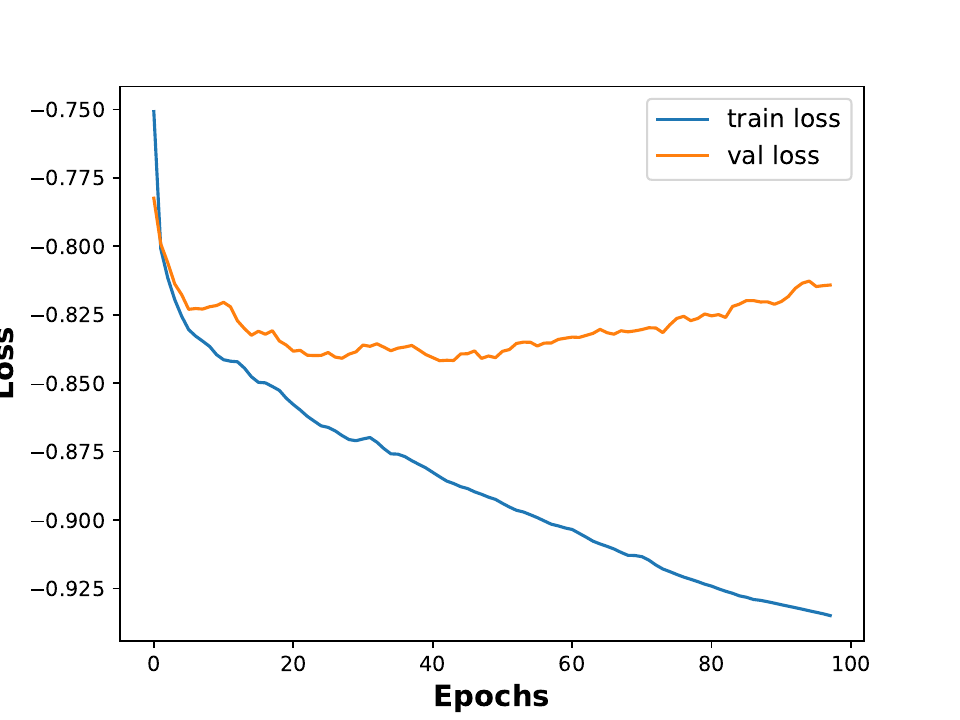}}

    \caption{We visualize the extracted features using t-SNE and show the corresponding train and validation loss for the LS-only, Mixed, and MMSE-only methods. The Mixed method achieves the highest average NMI and the shortest average training time compared to MMSE-only, while LS-only performs the worst.}
    \label{cl results}
\end{figure*}

\subsection{Contrastive Learning Results} 
In the CL training, the model is trained for 100 epochs with a batch size of 128. The learning rate is set to \(10^{-3}\) and drops to \(10^{-4}\) gradually using cosine decay schedule. The source dataset is split into 90\% for training and 10\% for validation, with a fixed random seed. The data augmentation process begins with adding AWGN, where the SNR is randomly selected from the range [10, 20] dB. Following this, LS and MMSE channel estimations are performed according to the selected SNR values. Finally, a block-wise masking operation is applied with a 10\% masking ratio to further enhance the model's robustness.

The CL loss and extracted features are shown in Fig. \ref{cl results}. Features are visualized via t-SNE~\cite{van2008visualizing}, and the performance is evaluated using the normalized mutual information (NMI) coefficient. A higher NMI indicates that the extracted features better capture the structure of the data, leading to more accurate clustering and better generalization of the model. The visualization shows the results after applying LS estimation equalized signals for the LS-only method and Mixed method, and MMSE estimation equalized signals for the MMSE-only method. It can be observed that the Mixed method performs better and achieves a higher average NMI coefficient. Additionally, compared to other methods, the Mixed method does not exhibit overfitting and produces a more separable feature space.

\subsection{Fine-Tuning Results}
To evaluate our method's performance in cross-scenario applications, we generate a test dataset using a new channel. After CL training, the model is fine-tuned using a few labeled data. The classification head for the fine-tune is shown in Fig. \ref{fine-tune-network}. The best NMI model is selected as the backbone, which is performed with a batch size of 10 for the fine-tuned dataloader and 512 for the validation dataloader. The test dataset is divided into 1\% fine-tune dataset (10 per device) and 99\% as a validation dataset. The random seed is set to ensure consistency. Early stop is used if validation accuracy does not improve in 30 epochs. The learning rate is set to \(10^{-3}\). We set the learning rate for the CNN layers to be 1/100 of the current learning rate, while the classification head uses the current one. During the fine-tuning stage, we also applied the AWGN data augmentation and combined LS and MMSE channel estimations to expand the dataset. On the validation set, we use a hybrid combination of LS and MMSE with equal weights for the final decision.

\begin{figure}
    \centering
    \includegraphics[width=3.3in]{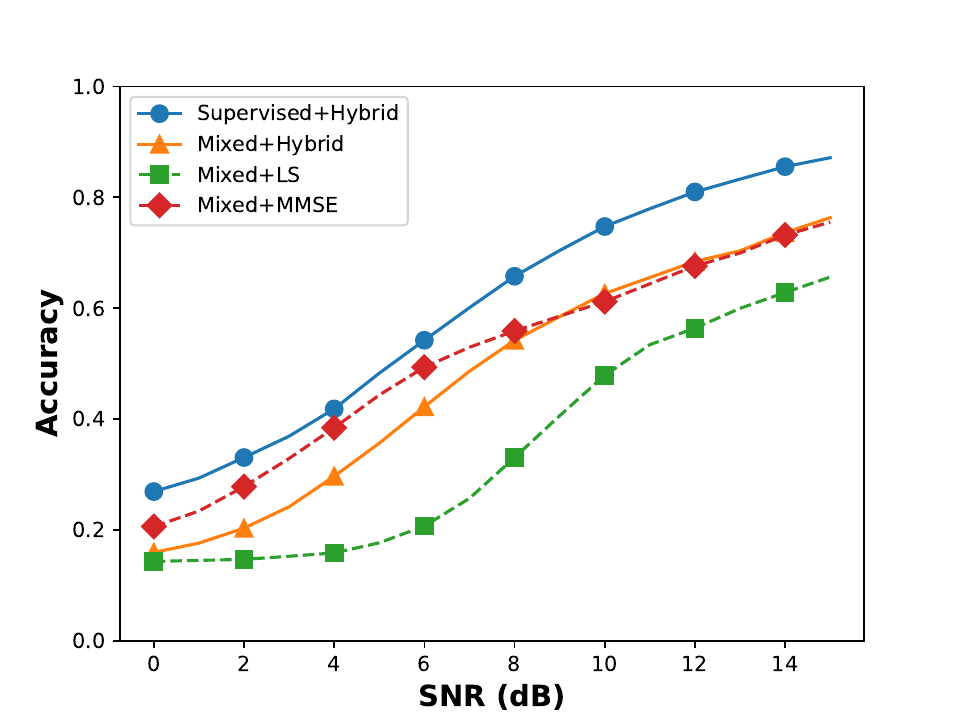}
    \caption{The SNR ranges from 0 to 15 dB, with a supervised model used as the baseline for comparison.}
    \label{finetune_SNR}
\end{figure}

The fine-tuned results are shown in TABLE \ref{finetunetable}. It can be seen that LS-only fails to work while MMSE-only performs slightly better. The Mixed method achieves satisfactory accuracy, demonstrating the effectiveness of our proposed method. The difference between supervised training (70\% labeled data) and our method is only 7\%, indicating that our method is highly competitive. The SNR performance is shown in Fig. \ref{finetune_SNR}, where the Mixed pre-trained model's performance is evaluated under lower SNR conditions. The results show that combination helps dataset expansion, but MMSE shows better performance due to its superior noise suppression capabilities.

\begin{table}[!t]
    \caption{Fine-Tune Accuracy for Validation Set in \(20dB\)}
    \label{finetunetable}
    \centering
    \begin{tabular}{c|cccc}
    \hline
    \textbf{}       & \textbf{LS-only}  &  \textbf{Mixed} & \textbf{MMSE-only} & \textbf{Supervised}     \\ \hline
    \textbf{Acc.}   & 14\%              &  82\%         & 44\%               & 89\%                    \\ \hline           
    \end{tabular}
\end{table}

\section{Conclusion}
This paper proposes a contrastive learning method using residual channels for data augmentation, with LS and MMSE estimations followed by equalization. The SimSiam network learns robust fingerprint features, achieving performance comparable to supervised learning with just 1\% of labeled data. The equalization removes residual channel interference without additional process, ensuring high efficiency. The lightweight model is computationally efficient, making it ideal for real-time tasks in wireless communication.

\bibliographystyle{IEEEtran}
\bibliography{ref.bib}
\end{document}